\documentclass[smallcondensed]{svjour3}

\usepackage[english]{babel}

\usepackage{amsmath}

\usepackage{graphicx}

\usepackage{amssymb}

\usepackage{mathptmx}       
\usepackage{helvet}         
\usepackage{courier}        
\usepackage{type1cm}        
\usepackage{makeidx}         
\usepackage{graphicx}        
\usepackage[bottom]{footmisc}
\usepackage{epsfig}
\usepackage{float}
\usepackage{bm}
\usepackage{mathrsfs}
\usepackage{amsfonts}
\usepackage[T1]{fontenc} 
\usepackage{lmodern,aecompl}
\usepackage[latin1]{inputenc}
\usepackage{latexsym}
\usepackage{verbatim}
\usepackage{subfigure}
\usepackage[usenames,dvipsnames]{color}

\journalname{Celest Mech Dyn Astron}

\begin{document}
\title{Phase correlations in chaotic dynamics\\A Shannon entropy measure}

\titlerunning{Phase correlations in chaotic dynamics}

\author{P.~M.~Cincotta and C.~M.~Giordano}

\authorrunning{P.~M.~Cincotta and C.~M.~Giordano}

\institute{Grupo de Caos en Sistemas Hamiltonianos\\ 
           Facultad de Ciencias Aston\'omicas y Geof\'{\i}sicas, 
           Universidad Nacional de La Plata and\\ 
           Instituto de Astrof\'{\i}sica de La Plata (CONICET-UNLP),
           Argentina\\ 
           \email{pmc@fcaglp.unlp.edu.ar}, \email{giordano@fcaglp.unlp.edu.ar}}

\date{Received: date / Accepted: date}

\maketitle

\begin{abstract}
In the present work we investigate phase correlations by recourse to the Shannon entropy. 
Using theoretical arguments we show that the entropy  provides an accurate measure of phase 
correlations in any dynamical system, in particular when dealing with a chaotic diffusion
process. We apply this approach to different low dimensional maps in order to
show that indeed the entropy is very sensitive to the presence of correlations among the successive values
of angular variables, even when it is weak. Later on, we apply this approach to
unveil strong correlations in the time evolution of the phases involved in the Arnold's Hamiltonian that
lead to anomalous diffusion, particularly when the perturbation parameters are comparatively large.
The obtained results allow us to discuss the validity of several approximations and assumptions usually
introduced to derive a local diffusion coefficient in multidimensional near--integrable Hamiltonian systems,
{in particular the so-called reduced stochasticity approximation.}

\end{abstract}

\section{Introduction}
\label{intro}

Chaotic diffusion in multidimensional near--integrable systems is a quite relevant dynamical 
process observed in different fields of astronomy, physics and chemistry.  Particularly in dynamical astronomy, diffusion plays a major role in planetary
and galactic dynamics, as recently reviewed and discussed in \cite{MCB16}, \cite{CGMB17}, \cite{MGCG18} and references therein.
The characterization of this instability is still an open subject, phase correlations mainly due to stickiness, prevent
in general the free diffusion that exhibits for instance an ergodic system, where the variance of any phase space 
variable, $\sigma^2$, increases linearly with time. This type of behavior is known in the literature as normal diffusion. 

As shown in several works, it seems that in the limit of {weak chaos\footnote{By weak chaos we mean
the dynamical state when the unstable chaotic motion is mostly
confined to the narrow stochastic layers around resonances.}}, the diffusion that proceeds along the chaotic layer of a single resonance 
(or along different layers of the resonance web) over very large motion times might be approximated
by a normal process (see for example 
\cite{LGF03}, \cite{FGL05}, {\cite{GLF05}}, \cite{FLG06}, \cite{LFG08}, \cite{EH13} and \cite{CEGM14}),  
leading then to a slow drift of the unperturbed integrals of motion. 
On the other hand, in the strongly chaotic scenario and
relatively short (or physical) motion times, diffusion could deviate significantly from the normal regime as it was presented in \cite{CGMB17}. This anomalous behavior 
of the diffusion imposes  serious limitations on the derivation of a diffusion coefficient. Indeed, within
this approach to the stability problem, the diffusion coefficient is introduced as the constant rate
at which the variance evolves with time.

Let us mention that anomalous diffusion has been observed also in many low dimensional dynamical
systems. In for instance \cite{Z94},\cite{ZA95}, \cite{ZEN97},
\cite{ZN97}, \cite{ZE00} and \cite{Z02}  as well  
as in \cite{KBZS90}, \cite{KZS93}, \cite{KK04}, \cite{SGM17}, \cite{V08}, \cite{MSV14}, a different
approach to diffusion is considered through a transport process formulation, where a generalization
of the normal regime is introduced, i.e.
the variance scales as $\sigma^2 = D t^b$ where the value of the exponent $b$ determines
the regime of anomalous transport, namely super--diffusion if $b > 1$ or sub--diffusion when $b < 1$, and 
the diffusion coefficient $D$ is thus defined. 

In \cite{GC18} we present an alternative way to measure both, the extent and the time rate of 
chaotic diffusion by means of the Shannon entropy (see \cite{SW49}). A relevant aspect of this approach is 
that the time evolution of the entropy is independent of the transport process that takes place 
in phase space and therefore a measure of the diffusion rate could be derived independently of the power law the variance satisfies. 

In any case, as we have already mentioned, phase correlations are responsible for anomalous diffusion.
The analysis of phase correlations is a quite difficult problem and thus restrictive assumptions are adopted in order to obtain analytical estimates for local diffusion coefficients
(see for instance \cite{Ch79}, \cite{C02}, \cite{CEGM14}). Therefore, herein we face with the problem 
of phase correlations in order to get some insight on the time evolution of chaotic diffusion.

While the time--correlation function (see for instance \cite{D87},
\cite{R98}) seems to be an appropriate tool to quantify phase correlations, its computation 
is not an easy task (as shown for example in \cite{BH70}) 
and  its numerical implementation is in fact quite expensive. Thus, in the present effort, 
also by recourse to the Shannon entropy, we provide a very efficient and simple way to
{clearly detect  phase correlations.}

In Section~\ref{theory} we outline the underneath theory of the Shannon entropy in a similar fashion
as that given in \cite{GC18}, while in Section~\ref{experiments} profuse numerical experiments are presented
in several well known maps to finally investigate phase correlations in the Arnold's Hamiltonian
\cite{A64}, the very same system studied in \cite{CGMB17} and \cite{GC18}. In this direction,
we discuss the limitations of the assumptions behind the analytical derivation
of expressions for a local diffusion coefficient in Chirikov's  formulation
of Arnold Diffusion \cite{Ch79}.

\section{The Shannon entropy}\label{theory}

As we have already discussed, it is not a simple task to provide a measure of 
correlations among successive values of a dynamical variable, in particular at large times.
Thus, in the present section we propose the Shannon Entropy as an efficient tool to 
measure correlations.

Theoretical background 
on the Shannon entropy can be found in \cite{SW49} however, different approaches to
the entropy in dynamical systems are presented in, for instance,
\cite{K67}, \cite{AA89} and \cite{W78}. Applications to time series analysis
can be found in \cite{CHMNV99} and \cite{C99}. A recent approach concerning the
use of the Shannon entropy to measure chaotic diffusion and its time rate is
addressed in \cite{GC18}.

Following for instance \cite{AA89} and \cite{L14}, consider the function $Z$ 
defined in $[0,1]$ as:
\begin{align}
Z(x)= \begin{cases}
-x\,\mathrm{ln}x,\qquad & x \in (0,1]\\\\ 
0 \qquad & x=0.
\label{zeta}
\end{cases}
\end{align}
It is evident that $Z(x)\ge 0$ and  
$Z^{\prime\prime} < 0$. Consider now an open domain $B\in\mathbb{R}^n$ and let
\begin{equation}
\alpha\,=\,\{a_i;\,i=1,\cdots,q\}
\label{part}
\end{equation}
be a partition of $B$, 
let us say a collection of $q$ $n$-dimensional cells that   
completely cover $B$.
The elements $a_i$  are assumed to be both measurable and disjoint.
{We consider}
$n = 1$ and $B = \mathbb{S}^1$ or the unit interval $(0,1)$ with opposite sides
identified.  

Let $x_i=x(t_i)$ be the phase variable of a given map $M$ (or Hamiltonian flow) such that 
$(y_n,x_n)\,\to\,(y_{n+1},x_{n+1})$, with $y\in G\subset \mathbb{R}$ some action variable~\footnote{In fact, the dynamical system could involve more than one action and phase.}.

For any finite trajectory $\gamma=\{(y_i,x_i)\in G\times B, i=1,\dots,N\}$ of $M$, let us define
$\gamma_x =\{x_i\in B, i=1,\dots,N\}\subset\gamma$, then a probability density on $B$ can be 
defined as
\begin{eqnarray}
	\rho({u},\gamma_x),=\,\frac{1}{N}
\sum_{i=1}^{N} \delta({u}-{{x}}_{i}),
\label{dens}
\end{eqnarray}
where $\delta$ denotes the {delta function}. It is clear that
\begin{eqnarray}
\int_B \; \rho({u},\gamma_x)\,d{u} \,=\,1,
\end{eqnarray}
and the measure $\mu(a_i)$ results
\begin{eqnarray}
\mu(a_i(\gamma_x))\,=\,\int_{a_i}\; \rho({u},\gamma_x)\,d{u}.
	\label{mu}
\end{eqnarray}
Then,  for the partition $\alpha$ the entropy
for $\gamma_x$ is defined as
\begin{eqnarray}
S(\gamma_x,\alpha)\,=\,\sum_{i=1}^{q}\,Z\left(\,\mu(a_i(\gamma_x)) \,\right)\,=\,
               - \sum_{i=1}^{q}\,\mu(a_i(\gamma_x))\,\ln(\mu(a_i(\gamma_x))).
\label{entropy}
\end{eqnarray}

Notice that for a given partition and trajectory, the entropy 
is bounded. Indeed, for the partition (\ref{part}) 
it is $0\le S(\gamma_x,\alpha)\le \ln q$, for any $\gamma_x$. 
The minimum corresponds to the situation in which the $x_i$  are restricted to a 
single element of 
$\alpha$, say the $j$--th element, which would correspond to the extreme
case of full correlation of the phase values at all times, such as for example at a first order 
resonance or one-period fixed point of $M$ ($x_{n+1}=x_n,\forall n$). 
In such a case it is $\mu(a_j)=1, \mu(a_i)=0, \forall i\ne j$, leading to $S = 0$. 
On the other hand, the maximum value, 
$S=\ln q$, would be obtained when the $q$ elements of the partition 
had an equal measure, that is $\mu(a_j)=1/q$, which corresponds to the situation in which 
the $x$ values are dense and uniformly distributed in  $B$. Therefore, in the case of
a non-integrable system, the entropy seems to be a natural measure of correlations.

Let $n_k$ be the number of phase values in the cell $a_k$,
then $\mu(a_k)=n_k/N$.  From the normalization condition,
$\sum_{k=1}^q\mu(a_k)=1,$ it follows that $\sum_{k=1}^qn_k = N.$
Therefore, the entropy given in (\ref{entropy}) reduces to
\begin{equation}
	S(\gamma_x,\alpha) = \ln N -\frac{1}{N}\sum_{k=1}^q n_k\ln n_k.
\label{entropyn}
\end{equation}

Take for instance  $\gamma_x^r =\{x_i=\theta^r_i\in B, i=1,\dots,N\}$ where the $\theta^r_i$ 
are random numbers and $N\gg 1$, 
then the $n_k$ obey a Poisson distribution:
$$P_{\lambda}(n) = \frac{\lambda^n}{n!}\mathrm{e}^{-\lambda},\qquad \lambda = \frac{N}{q},$$
where $\lambda$ is the mean value of the distribution as well as its variance.
Let us {assume $\lambda \gg 1$ and} take $n_k = \lambda + \xi_k$ with $|\xi_k|\ll \lambda$ ($n_k\in\mathbb{R}$), so
\begin{equation}
\sum_{k=1}^{q}\xi_k = 0,
	\label{xi}
\end{equation}
which clearly follows from the normalization condition. Then it is straightforward
to show that, up to second order in $\xi_k/\lambda$, it is 

\begin{equation}
	\sum_{k=1}^q n_k\ln n_k = N\ln N - N\ln q + \frac{1}{2\lambda}\sum_{k=1}^q\xi_k^2,
\label{entropyreduce}
\end{equation}
and  (\ref{entropyn}) reads
\begin{equation}
S(\gamma_x^r,\alpha) = \ln q -\frac{q}{2N^2}\sum_{k=1}^q\xi_k^2.
	\label{Saprox}
\end{equation}
Now on introducing the \emph{information} $\mathcal{I}$ as
\begin{equation}
\mathcal{I}(\gamma_x,\alpha) \equiv 1 - \frac{1}{\ln q} S(\gamma_x,\alpha),
\label{info}
\end{equation}
and estimating $\xi_k^2=\lambda$ we get,
in this particular limit of absence of correlations, that 
\begin{equation}
	\mathcal{I}(\gamma_x^r,\alpha) \approx \frac{q}{2N\ln q} > 0,
\label{I}
\end{equation}
which results independent of $\gamma_x^r$.
Since $N$ denotes the number of iterates of the map, we observe
that, for completely random motion, $\mathcal{I}$ would decrease with time as $t^{-1}$. 

On the other hand, from (\ref{entropy}) the time derivative of the entropy results
\begin{equation}
\frac{dS}{dt}(\gamma_x,\alpha)= -\sum_{k=1}^q\dot{\mu}(a_k)(1+\ln\mu(a_k)), 
\label{sdot0}
\end{equation}
and, for random motion, its theoretical estimate is 
\begin{equation}	
\frac{dS}{dt}(\gamma_x^r,\alpha)\approx \frac{q}{2N^2}\dot{N} > 0,
\label{sdot}
\end{equation}
which is seen to decrease as $t^{-2}$. 

The above results clearly state that,  for a random system, the information 
decreases as time increases, or alternatively, the entropy would
reach asymptotically  its maximum value; the system approaches the full 
mixing as $t^{-1}$.

Now let us consider $\gamma_x^s =\{x_i=\theta^s_i\in B, i=1,\dots,N\}$ with $N\gg 1$ and where $\theta^s_i$
are correlated phases, for instance  $\theta^s = \omega t$ with $\omega\in\mathbb{R}\setminus\mathbb{Q}$, which
represents the motion on a torus. Being $\omega$ irrational, the motion is ergodic on $\mathbb{S}^1$, thus
the distribution function of $n\in\mathbb{R}$ approaches $f(n_k) \approx \delta(n_k-\lambda)$, with $\lambda=N/q$.
Therefore, if we consider again $n_k = \lambda +\xi_k$ with $|\xi_k|\ll\lambda$, up to second order in 
$\xi_k/\lambda$, (\ref{Saprox}) holds.  The fluctuations 
should satisfy $|\xi_k|<1$, since every $n_k$ would
be the closest integer number to $\lambda$ which, in general, is a real quantity. Therefrom, setting $|\xi_k|\approx 1/2$, 
the information (\ref{I}) for a single quasiperiodic trajectory on the torus reduces to
\begin{equation}
\mathcal{I}(\gamma_x^s,\alpha)\approx \frac{q^2}{8N^2\ln q},
\label{Itorus}
\end{equation}    
and, in this particular case, the information would decrease as $t^{-2}$.
Notice that if we consider instead an ensemble of $n_p$ nearby trajectories, 
the estimate $|\xi_k|\approx n_p/2$ has to
be adopted (see below) and since $N = n_p\times t$, (\ref{Itorus}) becomes
\begin{equation}
\mathcal{I}(\gamma_x^s,\alpha)\approx \frac{q^2}{8\ln q}\frac{1}{t^2}.
\label{Intorus}
\end{equation}  

\begin{figure}[b!]
\begin{tabular}{cc}
\hspace{-5mm}\includegraphics[scale=0.50]{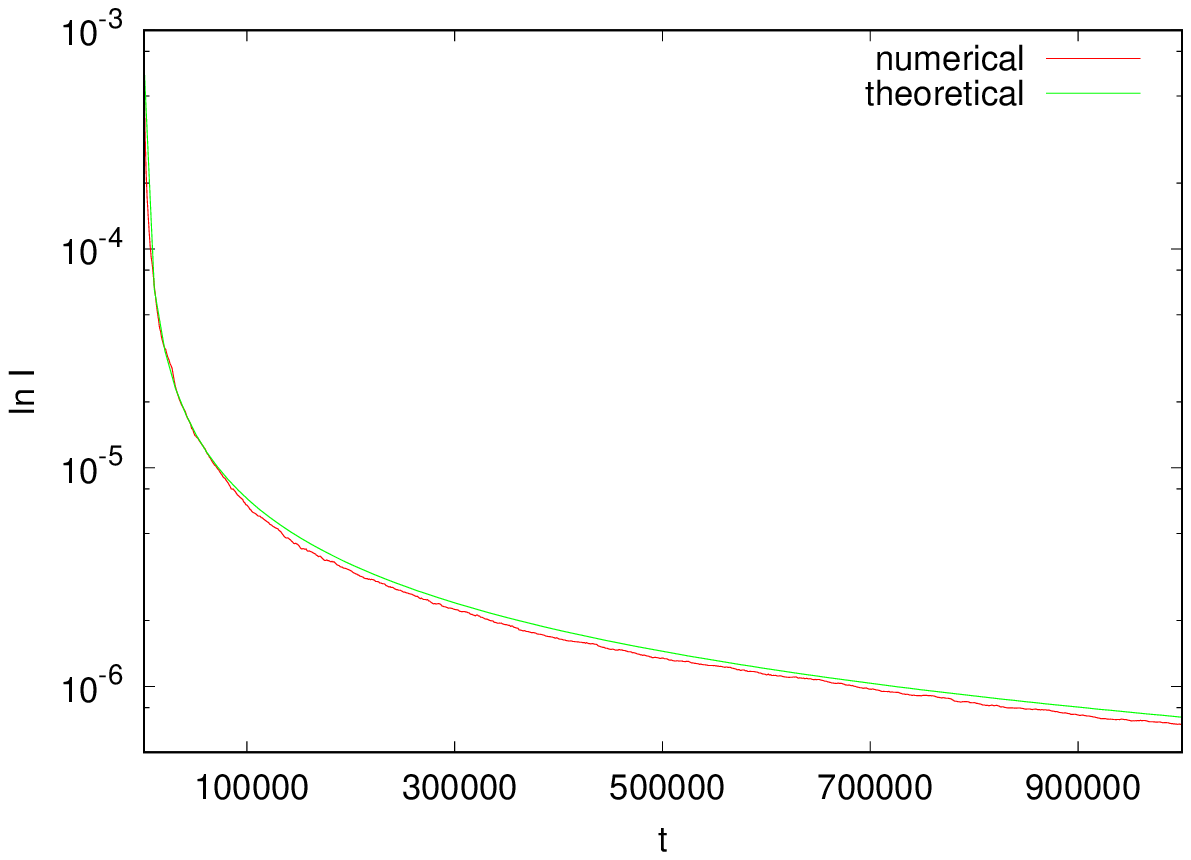}
\hspace{0.0mm}\includegraphics[scale=0.50]{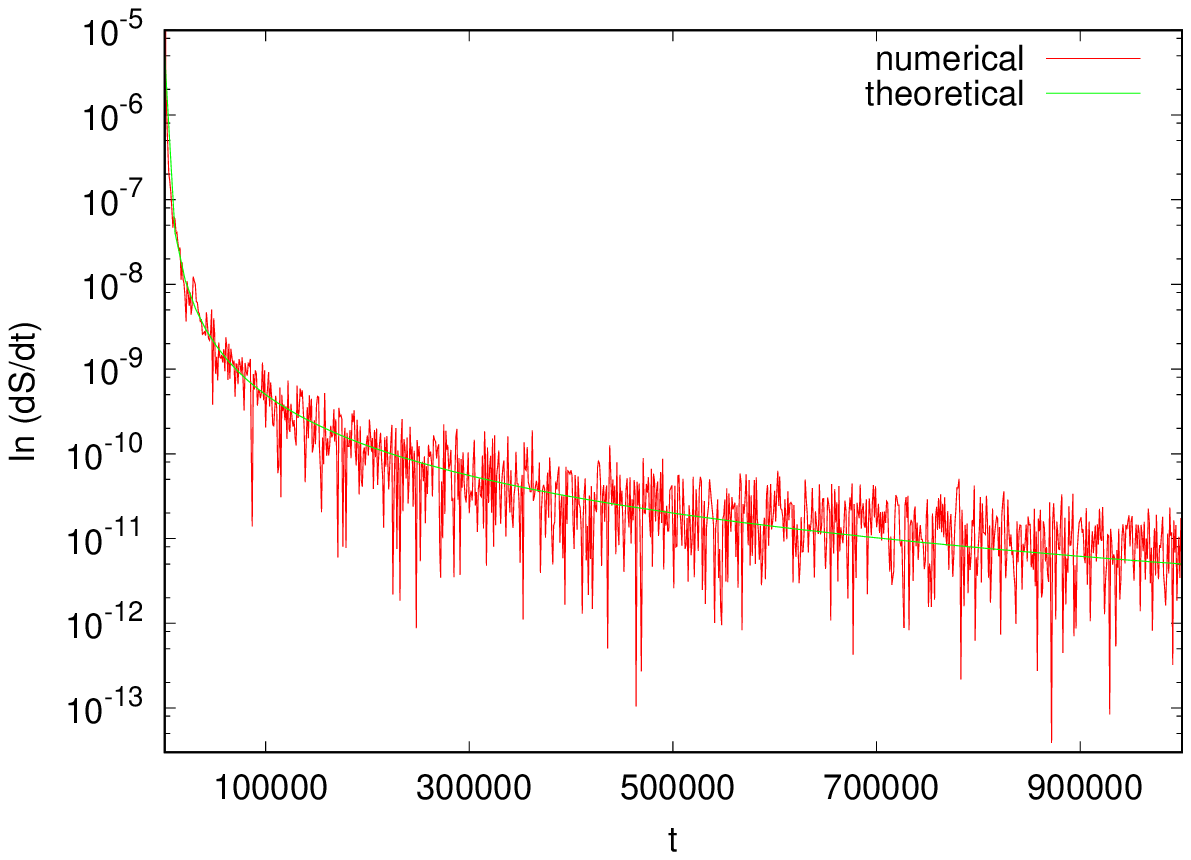}
\end{tabular}
\caption{Left panel: evolution of $\mathcal{I}$ for $n_p=100$ initial values of $x_i\in(0,1)$
evolving randomly (in red) and the expected theoretical trend given by (\ref{I}) with
$N = n_p\times t$ (in green). Right panel: the time derivative of the entropy computed numerically (in red) and
 the theoretical evolution given by  (\ref{sdot}) (in green), see text for details.}
\label{random}
\end{figure}

In order to check the analytical estimates given in (\ref{I}) and (\ref{sdot}), we pick up 
$n_p = 100$ initial values of the phases $x_i\in(0,1)$ and evolve each initial condition  
in a random way taking $x~\mathrm{mod}~1$ to finally  compute the information given by (\ref{info}) 
and the derivative of the entropy every $\Delta t = 10^3$. The information $\mathcal{I}$ for the ensemble  
corresponding to a partition $\alpha$ defined by $q=10^3$ cells on the unit interval, 
is computed for $N = n_p\times t$, where $t=10^3,\dots, 10^6$. 
The time derivative of the entropy can be evaluated, either from equation  
(\ref{sdot0}) or by the numerical derivative corresponding to the time evolution of (\ref{entropy}), also every $\Delta t = 10^3$. In fact, both results completely agree, though the latter procedure results  somewhat less noisy than the former,  so from now on we will restrict ourselves to show solely the numerical derivative of the entropy. 
Fig.~\ref{random} displays the results concerning the numerical simulation and the theoretical estimates for
both the information and the time derivative of $S$, revealing the consistency of the proposed analytical approach. 

Now, let us consider a similar example as the one above but for the 1D-map  
$x_{n+1} = x_n +\omega$
with $x~\mathrm{mod}~ 1$ and $\omega\in\mathbb{R}$. We have carried out several experiments, 
taking different values of $n_p$ ranging 
from $1$ to $2000$, ensembles of size $10^{-5}$ to $10^{-9}$ and varying both $\Delta t$ and $q$. 
The outcome values of $\mathcal{I}$ given by (\ref{info}) show to be only sensitive 
to the partition as (\ref{Intorus}) foresees, and 
to the $\omega$  value which could be either rational or irrational. For the particular case 
$\omega = l\in\mathbb{Z}$ we obtain  $\mathcal{I}\approx 1~(S \approx 0)$ despite of 
the size of the partition $q>1$. 

\begin{figure}[h!]
\sidecaption
\includegraphics[scale=0.6]{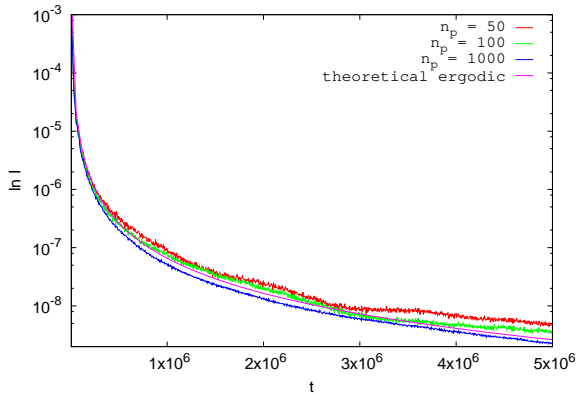}
\caption{Evolution of $\mathcal{I}$ for three different $n_p$ values,
	$|\omega-\sqrt{3}|\le 10^{-8}$ and  $q = 2000$ 
given by the map $x_{n+1} = x_n + \omega$ with $x_0=0$, and the expected theoretical trend for ergodicity
	given by (\ref{Intorus}) which is independent of $n_p$.}
\label{regular}
\end{figure}

In Fig.~\ref{regular} we present the representative behavior of $\mathcal{I}$ corresponding
to ergodic motion on a torus for different $n_p$ values in 
ensembles with $|\omega-\sqrt{3}|\le 10^{-8}$, $x_0 = 0$  
and $N = n_p\times t$, $t=5\times 10^3,\dots, 5\times10^6$, for a partition of the
unit interval of $q=2\times 10^3$ and $\Delta t = 5\times 10^3$. We also include in the figure the theoretical 
estimates given by (\ref{Intorus}).
Again, we notice a good agreement between the numerical and theoretical estimates for the information 
in the case of ergodic motion.  

The above  examples, though rather simple, allow us to assert that in a non-linear system the information, 
that measures the phase correlations, should decrease 
at most as $t^{-1}$, the limit taking place when the system becomes completely mixed. 
Strong phase correlations would provide larger values
of the information and $\mathcal{I}$ should scale with time as $\mathcal{I}\sim t^{\beta}, \beta > -1$. On the other hand,
if $\beta \to -2$,  the phase evolves as in an integrable system, the successive values of $x$  though fully
correlated are ergodic on the torus and so that the approximation
$\langle \sin^2 x\rangle = \langle\cos^2 x\rangle \approx t/2$ holds, though this result does not imply
that the phase values are completely uncorrelated.

\section{Numerical experiments}\label{experiments}
In this section we describe the numerical experiments regarding three different maps and provide evidence of how the information succeeds in yielding a measure of phase correlation in each case.

\subsection{The standard map}
This well known discrete area preserving 
dynamical system, introduced by Chirikov in \cite{Ch69}, \cite{Ch79} and 
largely studied in the
last four decades (see for instance \cite{MSV14} and references therein), 
$(I,\theta)\to(I',\theta'),\,$ 
is defined by
\begin{equation}
	I' = I + K\sin\theta,\qquad \theta' = \theta + I',
	\label{sm}
\end{equation}
with $ I\in\mathbb{R},\,\theta\,\,\mathrm{mod}\,\,2\pi$, and $K$ a free parameter. 
We numerically investigate this system which describes
the dynamics of a pendulum perturbed by an infinite set of periodic 
kicks of similar amplitude and frequency $\Omega = 1$. 
In fact we deal with the reduced standard map defined in $\mathbb{S}^1\times\mathbb{S}^1$ 
as a consequence of introducing the new variables $(p,x)$ that comply $I = 2\pi p,\, 
\theta = 2\pi x$, both $p\,$ and  $x\,\,$ being $\mathrm{mod}\,\, 1$, so that 
\begin{equation}
	p' =  p + k\sin 2\pi x,\qquad x' = x + p',\qquad k = \frac{K}{2\pi}.
	\label{rsm}
\end{equation}
It is well known that for $K < 1\,$ $\, (k < 1/2\pi)$ the motion is mostly stable, except near the thin chaotic
layers or homoclinic tangles around the integer resonance $p = 0$ of size $\sqrt{k}$, 
and the main fractional
resonances $p = 1/2$ and $p = 1/3, 2/3$, whose half-widths are $k$ and
$k^{3/2}$ respectively. 
For $K = K_c\approx 1$ the overlap
of resonances becomes relevant and this critical value separates stable from 
unstable motion in a broad sense. For $K > 4$ the centers of the integer resonances
become unstable and therefrom the motion is usually assumed to be almost completely uncorrelated. 

\begin{figure}[t!]
\begin{tabular}{cc}
\hspace{-5mm}\includegraphics[scale=0.50]{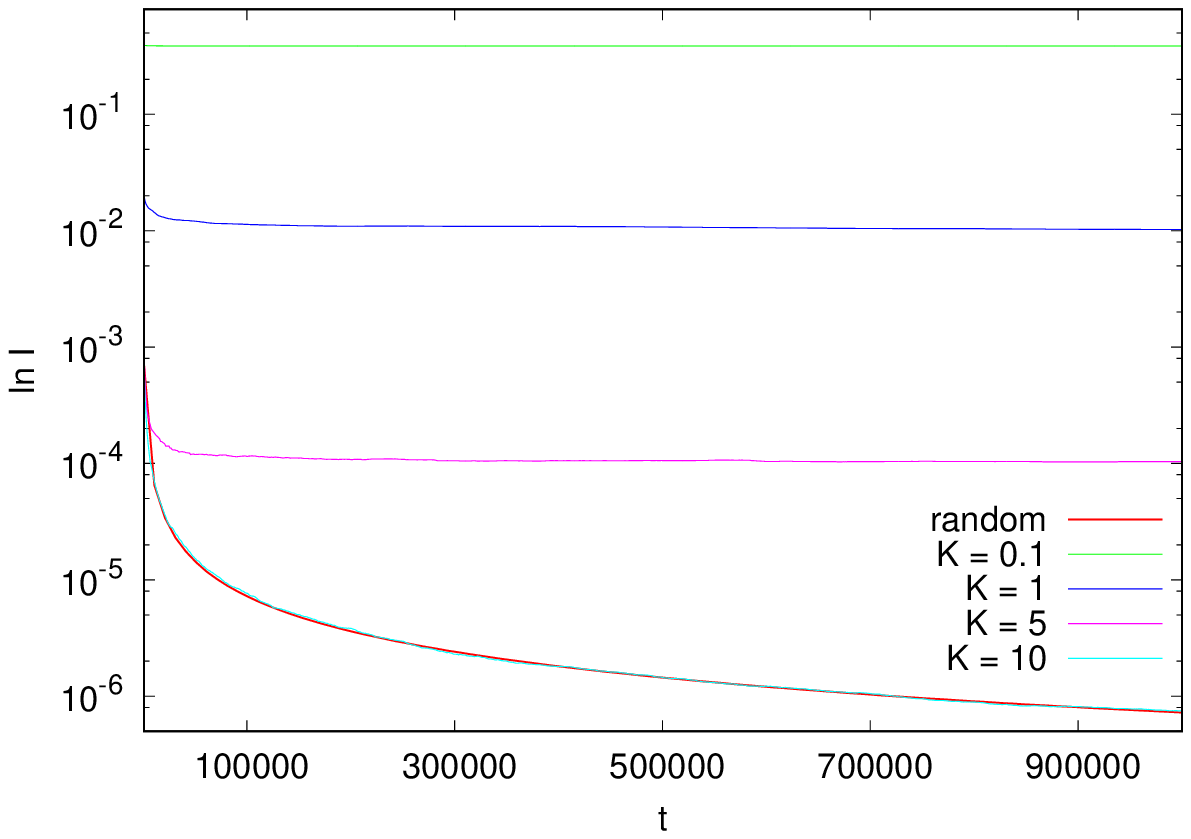}
\hspace{0.0mm}\includegraphics[scale=0.50]{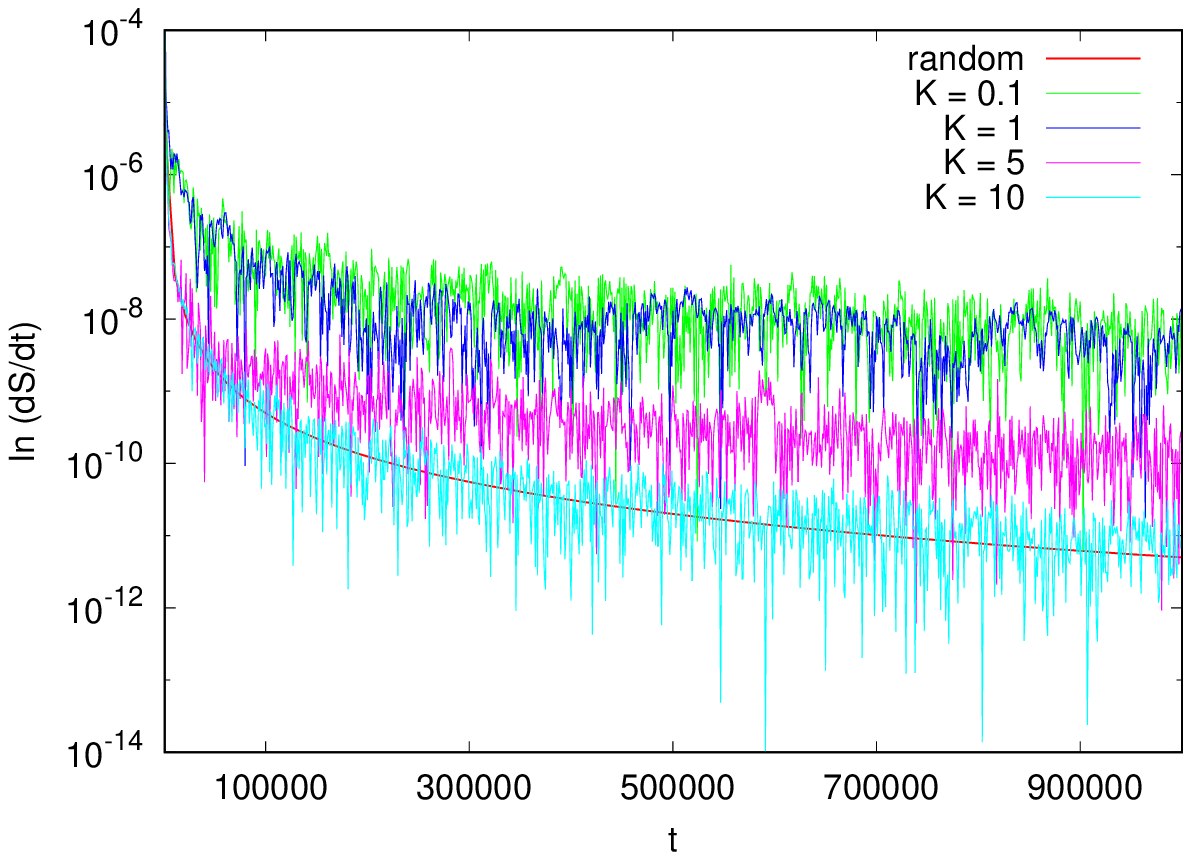}
\end{tabular}
\caption{(Left panel) Evolution of $\mathcal{I}$ for $n_p=100$ initial values of $x_i\in(0,1)$
	of the reduced standard map (\ref{rsm}) for different values of the parameter
	$K$ and the analytical expected behavior for random motion  given by (\ref{I}) displayed
	in red for $N = n_p\times t$. (Right panel) Similar to the plot at the left but 
	for the time derivative of the entropy computed numerically  and
        the theoretical expected evolution given by  (\ref{sdot}).}
\label{standard}
\end{figure}

We have performed several numerical experiments with values of $K$ in the interval $0\le K\le 100$ though  herein we only show  a few examples corresponding to $K = 0.1, 1, 5, 10$, for an ensemble of $n_p = 100$ random initial conditions with $p,x\approx 10^{-5}$. 
We have iterated the map (\ref{rsm}) up to $N = 10^6$, introducing a partition defined by $q = 10^3$, 
and we have computed the information (\ref{info}) and the derivative of the entropy (\ref{sdot0})
adopting a time interval of $\Delta = 10^3$. The results are displayed in Fig.~\ref{standard},
where we have also included the theoretical estimates (\ref{I}) and (\ref{sdot}) for  
$\mathcal{I}$ and $dS/dt$ respectively, which correspond to  the case
of random motion, for a partition defined by $q$ and $N = n_p\times t$.
The left panel shows the evolution of the information corresponding to 
different $K$ values.
Notice that the phase values appear to be correlated for the lower values of $K$; in fact, even for $K$ values as large as $1$ or $5$ the
successive values of the phases seem to be correlated for large times. A similar result is 
obtained for $K = 9$ (not shown in the figure), but for $K\ge 10$ the dynamics in $x$ looks like random,
the information decaying as $t^{-1}$.  
The right panel of Fig.~\ref{standard} corresponds to  the time derivative of the entropy. Besides the expected noisy
behavior, it turns out evident that $dS/dt$ decreases as $t^{-2}$ only for
the experiment corresponding to $K = 10$. These results suggest that 
 the usual assumption for the standard map,  
$$\langle(I(t)-I(0))^2\rangle = K\sum_{l=1}^t\langle\sin^2\theta(l)\rangle \approx \frac{K}{2}t,$$
is a plausible approximation only for $K\gtrsim 10$, which is in agreement with the results given  
in \cite{MSV14}.

\subsection{The whisker mapping}

Let us consider another canonical map, 
the whisker or separatrix mapping 
$(w,\tau)\to(w',\tau'), w\in\mathbb{R},\,\tau\,\,\mathrm{mod}\,\,2\pi$,
also introduced by Chirikov in \cite{Ch69},\cite{Ch79} as 
\begin{equation}
	w' = w + W\sin\tau,\qquad \tau' = \tau + \lambda \ln\left(\frac{32}{|w'|}\right),
	\label{wm}
\end{equation}
where $W,\lambda$ are free parameters. This map represents the motion close to the
separatrix of a non-linear resonance or pendulum subjected to a symmetric periodic 
perturbation; thus  $|W|\sim A_2(\lambda)\ll 1$ for $\lambda\gg 1,$  where
$A_2(\lambda)$ is the Melnikov-Arnold integral for $m = 2$. 
This simple system models the chaotic layer around resonances.
After
rescaling the a-dimensional energy $w$ by $\lambda W$ and defining $y = w/(\lambda W)$, the mapping (\ref{wm})
reduces to
\begin{equation}
	y' = y + \frac{1}{\lambda}\sin\tau,\qquad \tau' = \tau - \lambda \ln|y'| + G,
	\label{wmn}
\end{equation}
where $G=\lambda\ln(32/\lambda|W|)$ is a constant. It is well known that, after the
above rescaling, the chaotic layer
has a finite width, $|y|\lesssim 1$, whose central part, $|y|\lesssim 1/4$,
is highly chaotic, with no stability domains, while its external part
reveals a divided phase space with large stability islands. Therefore, the strong correlation between 
the successive values of the phase $\tau$ for $|y|\sim 1$ should be responsible for the
finite width of the layer. Roughly, near the edges of the layer it is $|y|\approx 1$ and thus
from the second relation in (\ref{wmn}) it is clear  that the structure of the external part of the layer
is dominated by $G$. Anyway, the study of its resonances, $y_{r_k} = (32/\lambda W)\exp(-2k\pi/\lambda)=
\exp((G-2k\pi)/\lambda)),$
$k\in\mathbb{Z}$, by means of  local standard maps clearly shows that the motion
for $|y|\approx 1$ or larger should be mostly regular.
Therefore it turns out fairly interesting to study phase correlations in this particular 
map by means of the Shannon entropy.

\begin{figure}[t!]
\begin{tabular}{cc}
\hspace{-5mm}\includegraphics[scale=0.50]{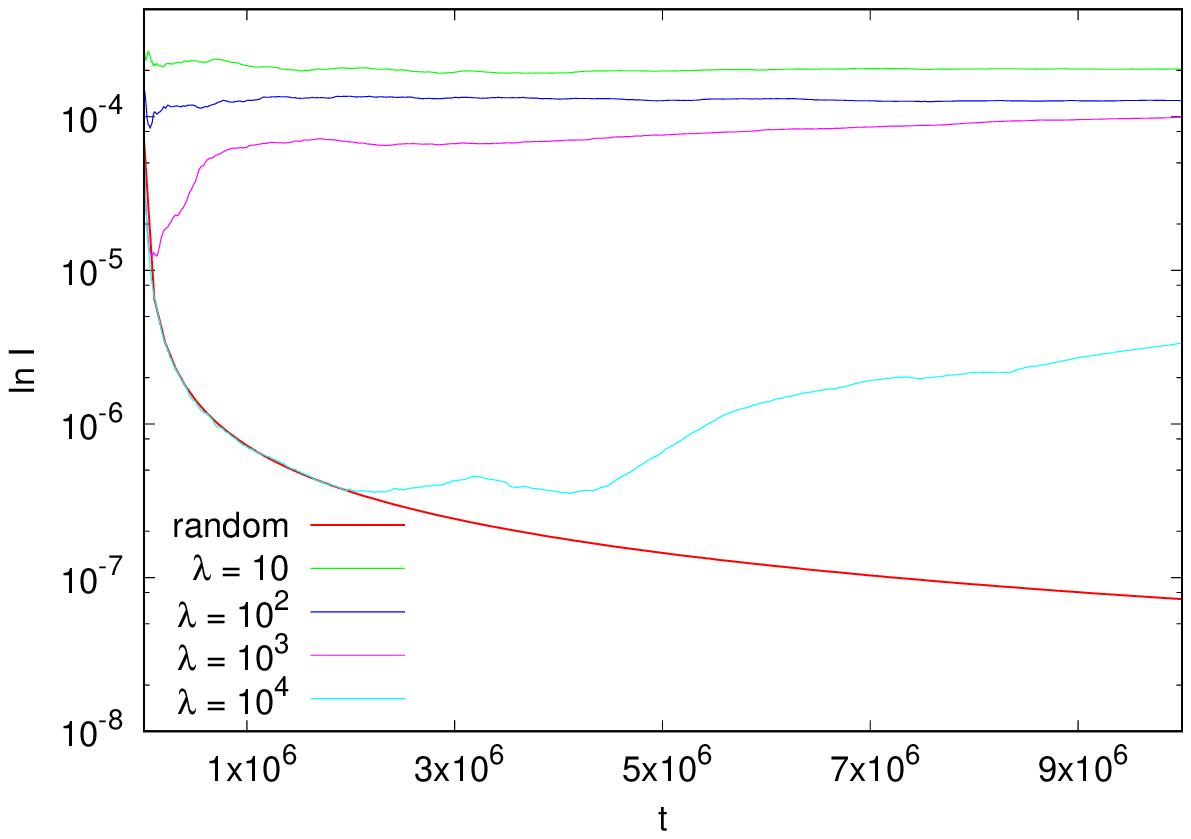}
\hspace{0.0mm}\includegraphics[scale=0.50]{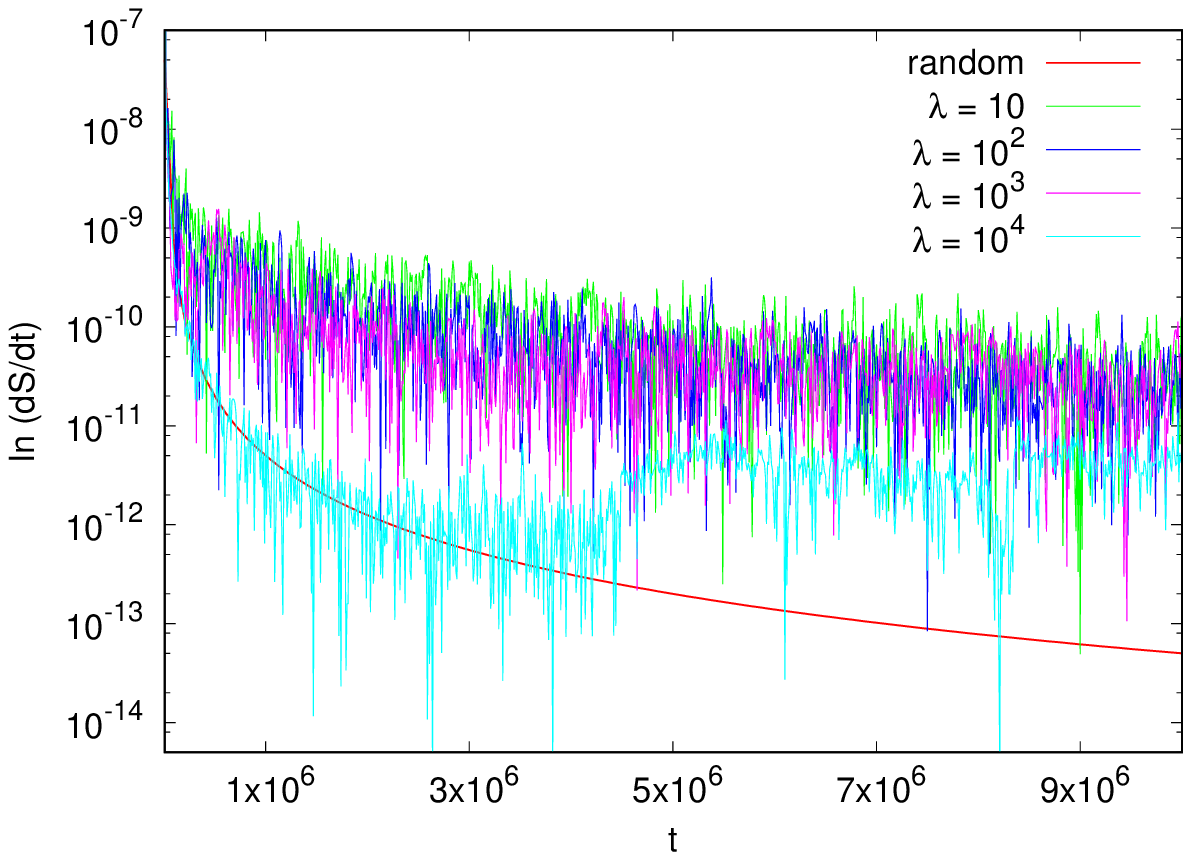}
\end{tabular}
\caption{(Left panel) Evolution of $\mathcal{I}$ for $n_p=100$ initial values of $x_i\in(0,1)$
        of the whisker mapping (\ref{wmn}) for $W = 10^{-6}$ and four different values of the parameter
        $\lambda$. The analytical expected behavior for random motion  given by (\ref{I}) is
        also included and displayed
        in red for $N = n_p\times t$. (Right panel) Similar to the plot at the left but
        for the time derivative of the entropy computed numerically  and
        the theoretical expected evolution given by  (\ref{sdot}).}
\label{whisker}
\end{figure}

Then, we have thoroughly investigated the map (\ref{wmn}) by numerical means taking different values
of the parameters $W$ and $\lambda$. Herein we include the results concerning a single value of $W$, namely  $W = 10^{-6}$, since
the map is almost independent of this scaling parameter, and  four different values of
$\lambda = 10, 10^2, 10^3, 10^4$. As before, we have considered an ensemble of $n_p = 100$ 
random initial conditions with $(y,\tau)\approx 10^{-5}$. We have iterated 
the map up to $N = 10^7$,
after introducing the normalized phase $x=\tau/2\pi\,\mathrm{mod~1}$, having taken a similar partition
of the unit interval, $q = 10^3$,  
and computed the information (\ref{info}) and the derivative of the entropy (\ref{sdot0})
with a sample time interval $\Delta = 10^4$. The results are displayed in Fig.~\ref{whisker},
where again we have included the theoretical estimates (\ref{I}) and (\ref{sdot}) corresponding to random motion for $\mathcal{I}$ and $dS/dt$ respectively. In any case, the information reveals that
at short times, while the trajectories are confined to the central part of the layer, the
values of the phase $\tau$ seem to be uncorrelated, but for large times instead, when the 
system has already reached the border of the layer, the correlations are quite evident, 
the information $\mathcal{I}$ increasing significantly. 
Notice that the time required to become a correlated system is seen to increase with $\lambda$,
as expected. The derivative of the entropy reveals the very same dynamical behavior.

\subsection{A generalized whisker mapping}
We devote this section to the numerical study of the  map that describes, accordingly to Chirikov, the diffusion 
across and along the chaotic layer of the resonance $\omega_1 = 0$ 
for the Arnold Hamiltonian (see \cite{A64}, \cite{Ch79}), which can be recast as 

\begin{equation}
	H(I_1,I_2, \theta_1,\theta_2,t;\varepsilon,\mu) = \frac{1}{2}(I_1^2+I_2^2)+{\varepsilon} (\cos\theta_1-1)[1+{\mu}(\sin\theta_2 + \cos t)],
	\label{arh}
\end{equation}
with $I_1, I_2\in\mathbb{R},\;\; \theta_1,\theta_2, t\in \mathbb{S}^1$, and $\mu\varepsilon\ll\varepsilon\ll 1$.
Let us note that (\ref{arh}) can be re-arranged  as

$$H(I_1,I_2, \theta_1,\theta_2,t;\varepsilon,\mu) = H_1(I_1,\theta_1;\varepsilon) + H_2(I_2) +
\mu V(\theta_1,\theta_2,t;\varepsilon),$$
where
$$H_1=\frac{1}{2}I_1^2+{\varepsilon} (\cos\theta_1-1),\qquad H_2=\frac{1}{2}I_2^2,\qquad V={\varepsilon}(\cos\theta_1-1)(\sin\theta_2 + \cos t),$$
and that the instantaneous change of $H$ and $H_2$ are given by
$$\dot{H} = \mu \frac{\partial V}{\partial t},\qquad \dot{H_2}=I_2\dot{I_2} =- \mu \omega_2\frac{\partial V}
{\partial\theta_2},$$
where $\omega_2=I_2\ne 0$ is assumed to be constant and irrational, then
\begin{eqnarray}
\dot{H}&=&\frac{\mu\varepsilon}{2}\left(2\sin t + \sin(\theta_1-t)-\sin(\theta_1+t)\right),\nonumber\\\label{doth}\\
\dot{H}_2&=&\frac{\mu\varepsilon}{2}\left(2\omega_2\cos\theta_2-\cos(\theta_2-\theta_1)-\cos(\theta_2+\theta_1)\right).
\nonumber
\end{eqnarray}
Thus, taking  $\theta_2(t) \approx \omega_2 t+\theta_2^0$, 
$\theta_1(t) = 4\arctan\left(\exp\left(\sqrt{\varepsilon}(t-t_0)\right)\right)$, i.e. the motion on the separatrix,
defined in such a way that $\theta_1=\pi$ for $t=t_0$, and
neglecting the free oscillatory terms
$\sin t$ and $\cos\theta_2$\,\footnote{See the discussion below.},  
the variations of $H$ and $H_2$ over a half--period of oscillation of $\theta_1$ result (see \cite{Ch79}
for more details)
$$\Delta H = \frac{\sqrt{\epsilon}\mu}{2}A_2\left(\frac{1}{\sqrt{\epsilon}}\right)\sin t^0,\qquad
\Delta H_2=\frac{\sqrt{\epsilon}\mu\omega_2}{2}A_2\left(\frac{\omega_2}{\sqrt{\epsilon}}\right)\cos\theta_2^0,$$
where $t^0$ is the value of the phase $t$ when the pendulum crosses the surface
$\theta_1=\pi$ where the stable equilibrium point lies, and $\theta_2^0 = \theta_2(t^0)$.

In Chirikov's formulation,
significant changes in $H$ are only possible due to large variations in
$H_2$, since $H_1$ is bounded to the exponentially small {(wrt $\mu$)} width of the chaotic layer. Moreover, defining
$D_H = \langle (H(t)-H(0))^2\rangle/t$ and $D_2 = \langle (H_2(t)-H_2(0))^2\rangle/t$, for large times  
it should be $D_H = D_2$ and thus 
\begin{equation}
\langle\sin^2 t^0\rangle = v^2\langle\cos^2\theta_2^0\rangle,
	\label{phases}
\end{equation}
with
\begin{equation}
v =\frac{\omega_2 A_2\left(\frac{\omega_2}{\sqrt{\varepsilon}}\right)}{A_2\left(\frac{1}{\sqrt{\varepsilon}}\right)}
	= \omega_2^2\,\,\frac{\sinh\left(\pi/\sqrt{\varepsilon}\right)}{\sinh\left(\omega_2\pi/\sqrt{\varepsilon}\right)}\,\,
	e^{\frac{(\omega_2-1)\pi}{2\sqrt{\varepsilon}}}\,
\approx \omega_2^2\, e^{\frac{(1-\omega_2)\pi}{2\sqrt{\varepsilon}}},
\label{v}
\end{equation}
where the approximation holds for $\varepsilon\ll 1$. The latter expression for the relative amplitude $v$
is significant only in the interval 
$|\omega_2-\omega_{2M}|<\sqrt{2\epsilon/\pi}$ where $\omega_{2M}=4\sqrt{\varepsilon}/\pi$ corresponds
to its maximum value, being the latter $v_M\approx (4/\pi e)^2\varepsilon\exp\left(\pi/(2\sqrt{\varepsilon})\right)$; 
and goes to zero in both limits
as $v\sim \omega_2^2,\, \omega_2\to 0$ and as 
$v\sim\exp\left(-\omega_2/\sqrt{\varepsilon}\right),\,\omega_2\to\infty$. Then, 
for $\omega_2 > 1$   it is $v \ll 1$. On the other hand, for $\omega_2$ in the narrow range
$\omega_2\approx \sqrt{\epsilon}\left(4/\pi \pm \sqrt{2/\pi}\right)$, it is
$v \gg 1$. 
Therefore from (\ref{phases})
for $\omega_2$ away from $0$ and $1$, it turns out that both phases could not be simultaneously random.

The associated map represents the finite variation of the energy in $H_1$ (pendulum model for the
resonance $\omega_1 = 0$) when the system is close to the separatrix of the resonance
$\omega_1=0$ and it can be computed as the difference
$\Delta H_1 = \Delta H - \Delta H_2$ after a half period of oscillation or a period of rotation 
of the pendulum and it has the form
$(y,\tau)\to(y',\tau')$,\, $y\in\mathbb{R},\, \tau\,\mathrm{mod~2\pi}$, where $y$ measures the energy changes in
$H_1$ relative to the separatrix energy ($H_1=0$), $\tau$ represents the successive values of $t^0$ and it 
reads
\begin{equation}
	y' = y + \sin \tau - v \cos(\omega_2 (\tau + \beta)),\qquad \tau' = \tau -\frac{1}{\sqrt{\varepsilon}}\ln|y'| + \eta,
\label{arm}
\end{equation}
where  $\eta$ is a  constant similar to $G$ introduced in the
whisker mapping, that depends on both $\mu$ and $\varepsilon$ but not on $\omega_2$,  
$\omega_2\beta$ being the initial value of $\theta_2^0$.

Chirikov proposed that the map (\ref{arm}) not only describes the thin chaotic layer around
the main or guiding resonance $\omega_1 = 0$ but the diffusion along the latter as well. For $\varepsilon \ll 1$
and $\omega_2 > 1\, (0 < \omega_2\sim 4\sqrt{\varepsilon}/\pi)$ since $v \ll 1\, (v \gg 1)$, in the right hand side  
of the first equation of (\ref{arm}) a clearly dominant term is present. Thus at this order, 
the map reduces to the whisker mapping, being the largest term (layer resonances in Chirikov's terminology) responsible for the properties
of the chaotic layer, such as its width and resonances' structure. Therefore, for large times, the successive 
values of the
phase involved in such a dominant term would be strongly correlated ($\tau$ for $\omega_2 \gg 1$ and $\theta_2$
for $\omega_2\sim 4\sqrt{\varepsilon}/\pi$). Note that for $v\ll 1\, (\omega_2\gg 1)$ the map becomes independent
of $\omega_2$ and thus, while $\tau$ values should present correlations, phase values $\theta_2^0=\omega_2\tau$
could evolve, in principle, nearly in a random or ergodic fashion when $\omega_2$ is irrational and large.

Since when the system is close to the separatrix of the resonance $\omega_1 = 0$, the variation in $H_1$ is given  by the largest
term in the map ($y$), the successive values of the smaller term in the first of (\ref{arm}) are 
assumed to be
nearly random, and this perturbation term ({or driving resonances accordingly Chirikov}) 
is responsible for a  nearly normal 
diffusion process along the resonance, i.e. large variations in $I_2$ such that
$\langle (H(t)-H(0))^2\rangle\propto t$ as in the standard map. In other words,
for $\omega_2\sim 4\sqrt{\varepsilon}/\pi<1$ it is assumed that $\langle\sin^2\tau\rangle = R/2$,
while for $\omega_2 >1, \langle\cos^2\theta_2^0\rangle = R/2$, where $R < 1$ is the so--called
reduction factor that takes into account some correlations among the successive phase values 
when the system is moving within the external part of the chaotic layer (see \cite{Ch79}, \cite{C02},
\cite{CEGM14}).

Let us remark that the derivation of the map (\ref{arm}) rests, among other simplifications, on
the following assumptions: (i) $\varepsilon\ll 1$ such that no overlapping
with other first order resonance takes place (as for instance among the resonances
$\omega_1=\pm 1$ and $\omega_1=0$); (ii) $\varepsilon\mu\ll 1$ such that 
we could consider  $\omega_2$ nearly constant and
(iii) $|I_2|\gg 2\sqrt{\varepsilon\mu}$, the width of the resonance $\omega_2=0$, 
such that $\cos\theta_2$ can be regarded as a purely oscillatory term. 

Therefore, the  expected behavior of both phases depending on $\omega_2$ encourages
the investigation of correlations in this map by means of the Shannon entropy.

We present the results of only one of the large set of experiments performed for  
$\varepsilon = 0.05$ and $\mu = 10^{-3}\varepsilon$ and eight different values
of $\omega_2^0 = \kappa\sqrt{3}, \kappa\in K$ where 
$$K = \{\kappa: \kappa = 0.06,\, 0.16,\, 0.25,\, 0.62,\, 1.75,\, 2.25,\, 11.0,\, 19.0\},$$ 
such 
that $0.10\lesssim \omega_2^0\lesssim 33$ and for $\kappa = 0.62, \omega_2\approx 1$. For these values of $\varepsilon$ 
and $\kappa\in K$, the relative amplitude $v$ (\ref{v}) ranges from $2\times 10^{-95}$ for
the largest $\kappa$ value up to $12$ for $\kappa=0.16$, which it is in agreement with the analytical 
estimates given above. The resonances' width is about $0.447$ for the resonance $\omega_1=0$ while it equals 
$0.003$ for the remainder ones, so that no overlap would be expected between the main resonance and
$\omega_1=\pm 1$.  For the smallest value of $\kappa\in K$, $\omega_2^0 \approx 0.104$ and it
would be plausible to assume that it is far from the resonance $\omega_2 = 0$ since its width
is quite small. Anyway, we numerically integrated the Hamiltonian (\ref{arh}) for the same set of parameters and
$n_p = 100$ random initial values of $\omega_2^0$ and $I_1^0$ on the chaotic layer of the 
resonance $\omega_1=0$ such that $I_1^0 = 2\sqrt{\varepsilon}$,
and we measured the variation of $\omega_2\approx I_2$ to find that the largest
value of $|\omega_2-\omega_2^0|\le 0.01$, which corresponds to $\kappa = 0.06$ after a total motion time $t=10^6$.  
It outcomes then that the approximations assumed to derive the map (\ref{arm}) 
are valid for the  adopted values of the parameters (see below). In Fig.~\ref{dif3d} we present a three dimensional
visualization of the diffusion for the smaller four values of $\kappa\in K$ for the section $|\theta_2|\le 10^{-5}$ where
we observe that at least up to $t=10^6$ the diffusion is almost irrelevant.
\begin{figure}[h!]
\sidecaption
\includegraphics[scale=0.6]{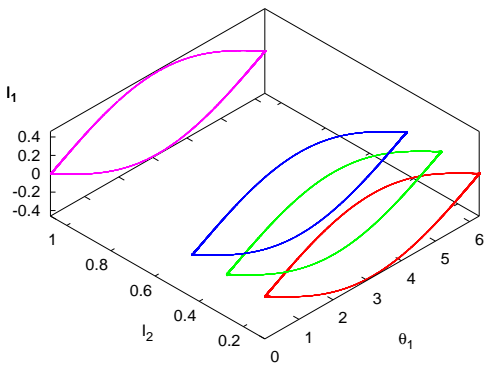}
\caption{Visualization of the diffusion along the chaotic layer of the resonance $\omega_1=0$ for four
values of $\kappa\in K$, $0.06$ in red, $0.16$ in green, $0.25$ in blue and $0.62$ in magenta.
}
\label{dif3d}
\end{figure}

\begin{figure}[t!]
\begin{tabular}{cc}
\hspace{-4mm}\includegraphics[scale=0.50]{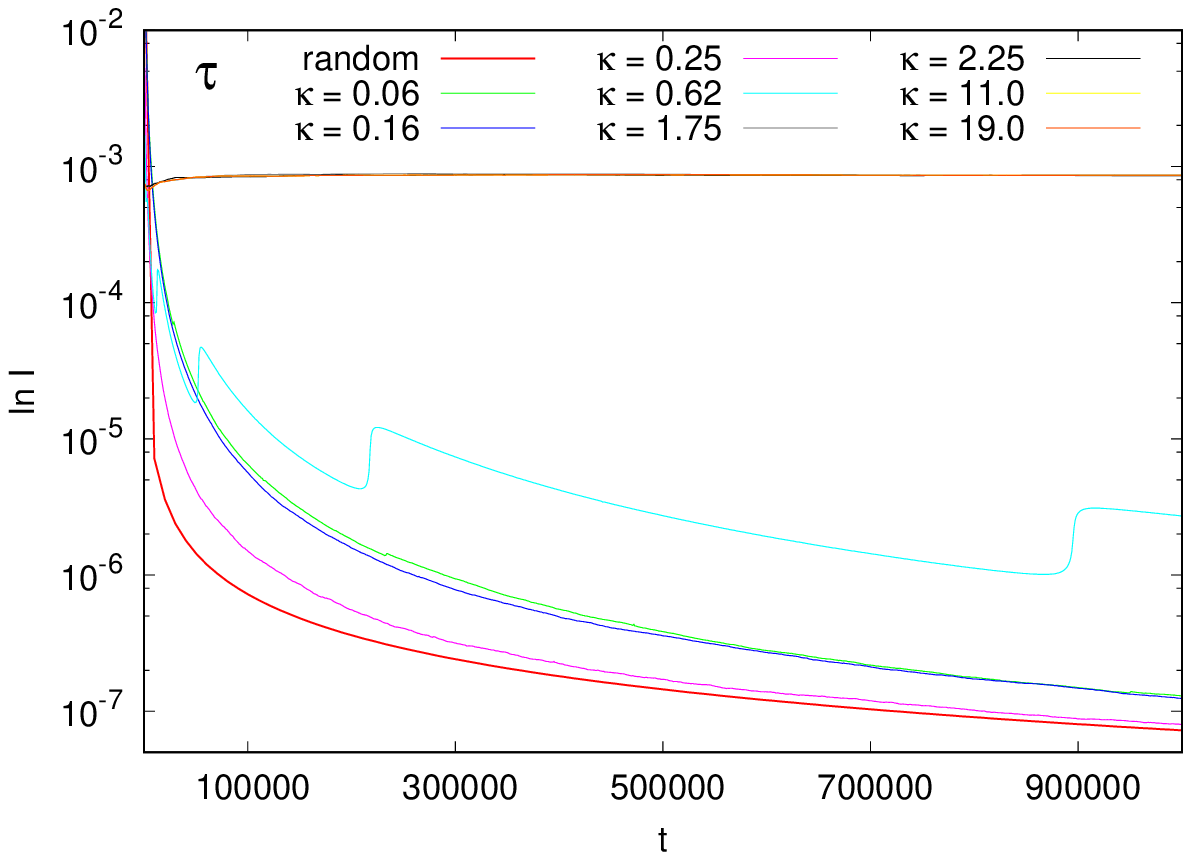}
\hspace{-2mm}\includegraphics[scale=0.50]{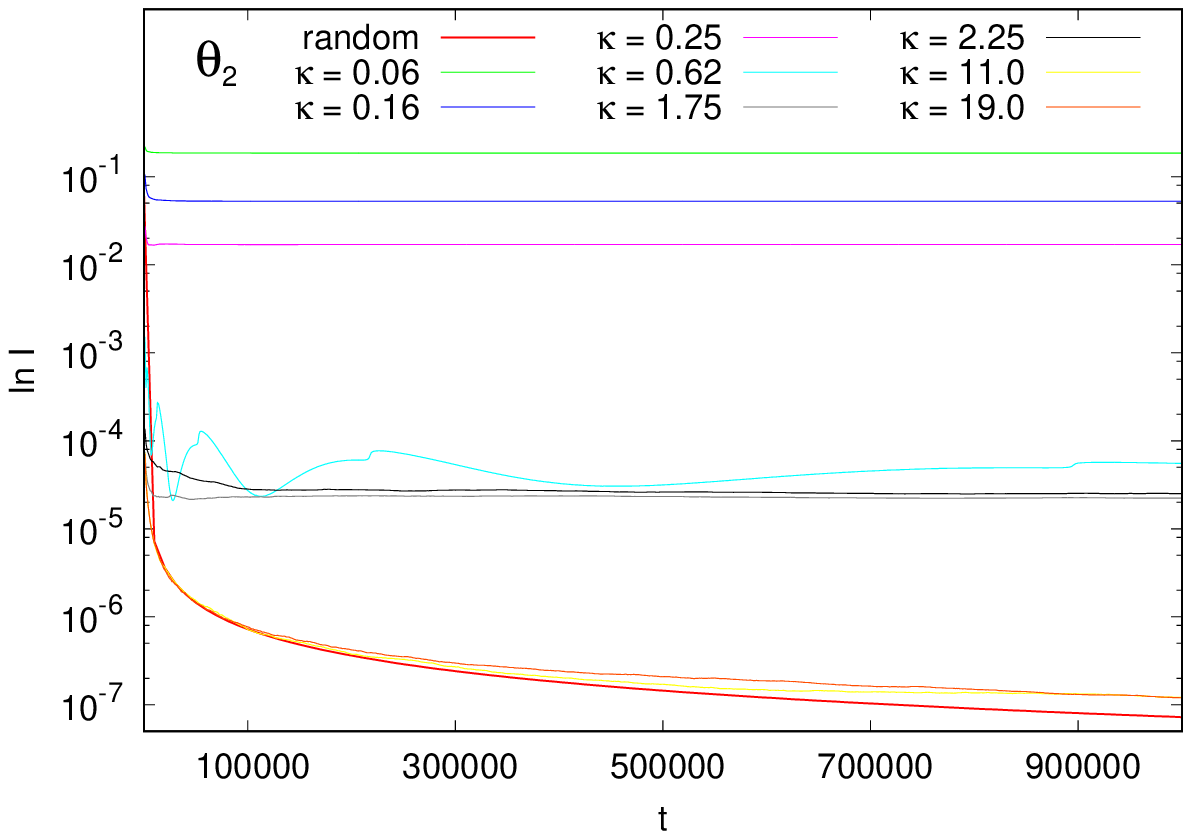}
\end{tabular}
\caption{(Left panel) Evolution of $\mathcal{I}$ for $n_p=1000$ initial values of $\tau\in(0,1)$
        for the map (\ref{arm}) for $\varepsilon = 0.05, \mu = 10^{-3}\varepsilon$ and eight
        values of $\omega_2^0 = \kappa\sqrt{3}$.
        The analytically expected behavior for random motion  given by (\ref{I}) is
        also included
        for $N = n_p\times t$. (Right panel) Similar to the plot at the left but for
        the phase $\theta_2\in(0,1)$.}
\label{arnold1}
\end{figure}

Again, we take for the map
an ensemble of $n_p = 1000$ random initial conditions with  $y\le 10^{-5}$, 
$|\omega_2 - \omega_2^0| \le 10^{-5}$, $\tau=10^{-5},\, \beta = 0$ and iterate the map (\ref{arm}) up to $N = 10^6$.
We normalize both phases $\tau$ and $\theta_2$ to the unit interval and compute the information
(\ref{info}) and the time derivative (\ref{sdot0})
after time intervals of length $\Delta = 10^3$ with $q=10^3$.
The results for the information are displayed in Fig.~\ref{arnold1}, where $\mathcal{I}$ as a function of time
is presented for both phases, $\tau,\theta_2,\,\mathrm{mod}(1)$, for the eight different values of $\omega_2^0$.
The left panel in Fig~\ref{arnold1} shows the results of the information concerning the phase $\tau$ while
the right panel does so for the phase $\theta_2$. We observe that
Chirikov's assumption might apply for $\omega_2 < 1$, where the phase $\tau$ behaves almost randomly though
some correlations are clearly present
while $\theta_2$ shows strongly correlated. On the other hand
for $\omega_2 > 1$, while it is evident that the successive values of $\tau$ are not sensitive to $\omega_2$ 
and highly correlated, only for large values of $\omega_2$  the phase $\theta_2$ seems to be uncorrelated,
when the amplitude $v$ is negligible. 
The results obtained from several experiments with different values of $\varepsilon$, $\mu$,
$\kappa$ and $N$ are quite similar when considering  $\mu\varepsilon\ll\epsilon\ll 1$ and the mentioned restrictions
to the location of the ensemble.
In sum, it seems that the conjecture that $\tau$ or $\theta_2$ are partially random, in the
sense that either $\langle \sin^2t^0 \rangle$ or $\langle \cos^2\theta_2^0\rangle$ (depending on  $\omega_2$'s value) 
grows in mean linearly with time, could be a fairly good  approximation provided that either $\omega_2\sim 4\sqrt{\varepsilon}/\pi<1$
or $\omega_2\gg 1$.  Nevertheless, some departures from the linear trend should be expected due to the presence of
correlations. 

\begin{figure}[h!]
\begin{tabular}{cc}
\hspace{-4mm}\includegraphics[scale=0.50]{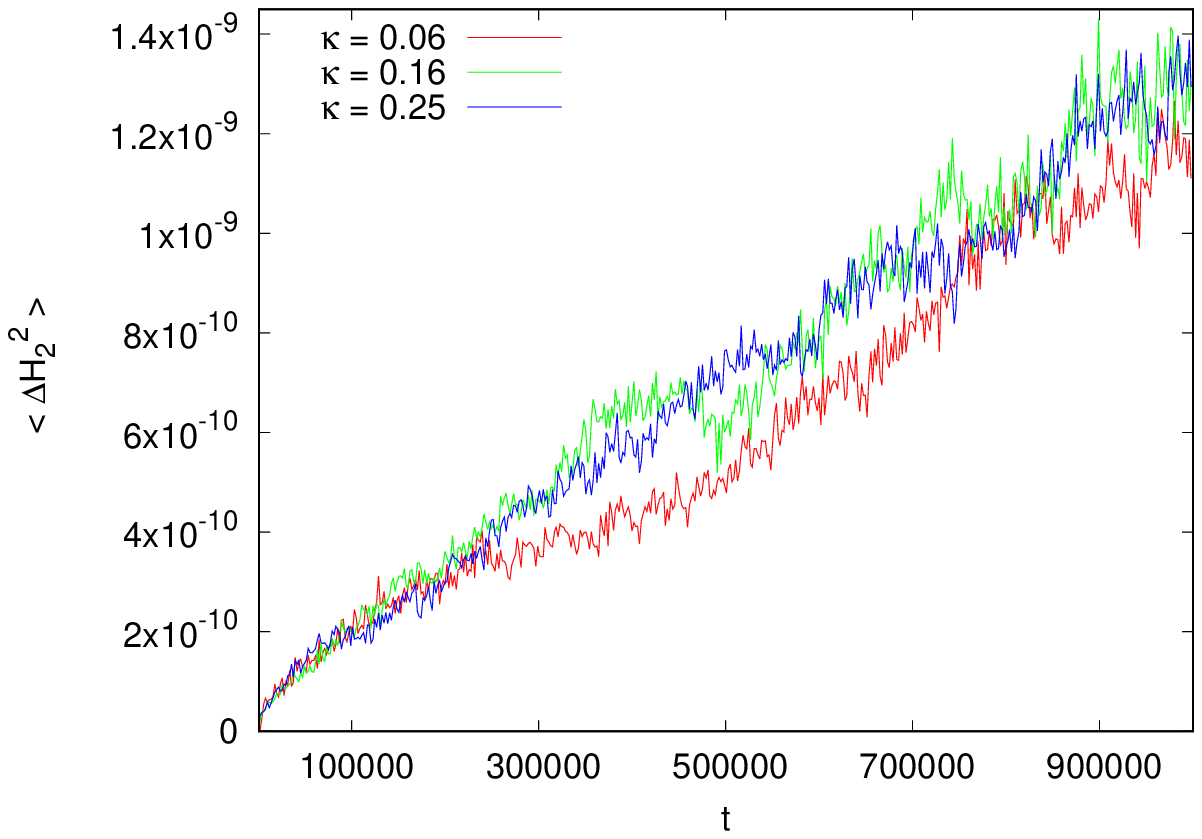}
\hspace{-2mm}\includegraphics[scale=0.50]{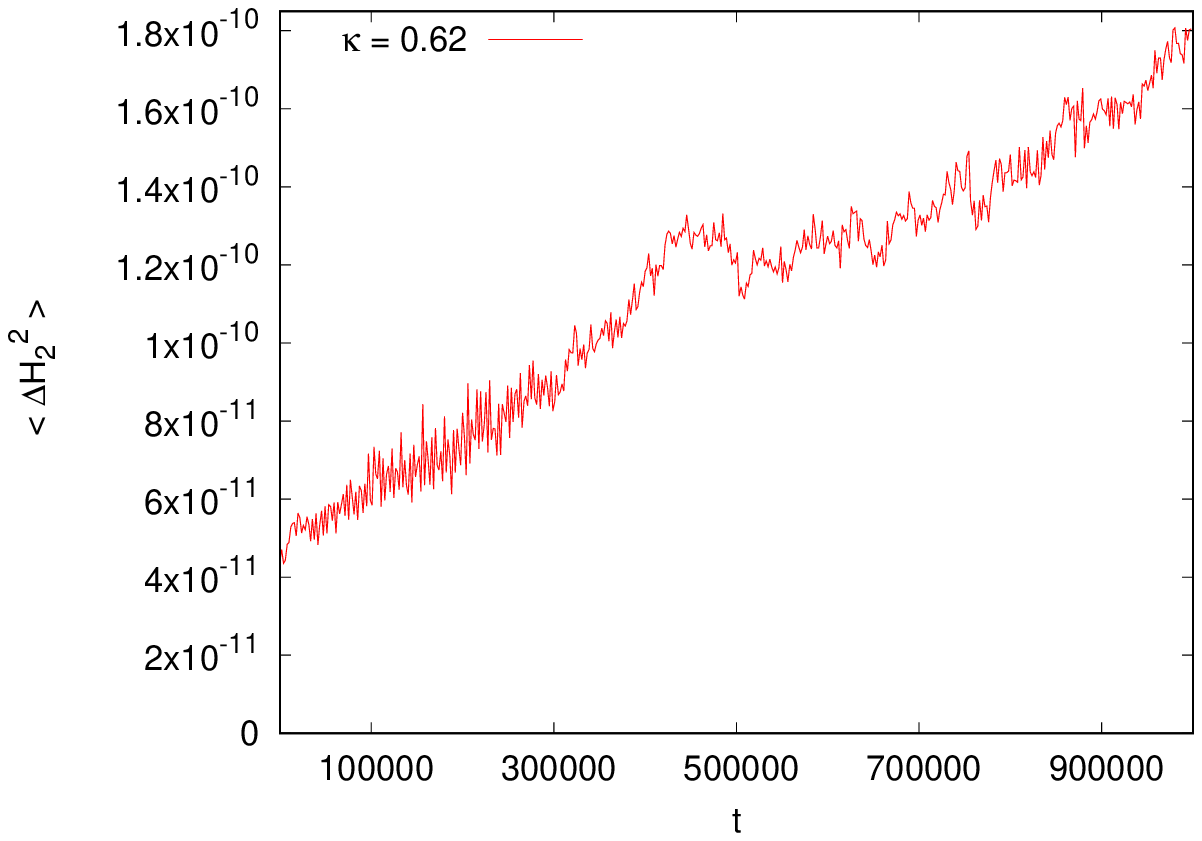}
\end{tabular}
	\caption{Evolution of $\langle \Delta H_2^2\rangle$ given in (\ref{ea}) for $\kappa = 0.06, 0.16, 0.25, 0.62$
	in different panels due to their different scale in the variance. The adopted time-step
	in this computation is  $\delta t = 5\times 10^{-4}$}
\label{var}
\end{figure}

Let us now consider the ensemble variance,
\begin{equation}
\langle \Delta H_2(t_j)^2\rangle=\frac{1}{n_p}\sum_{k=1}^{n_p}\left(H_2^{(k)}(t_j)-H_2^{(k)}(0)\right)^2,
\label{ea}
\end{equation}
where $t_j = t_0+j\delta t,  j\in\mathbb{Z}^+$, $\delta t\ll 1$~(see below) being the integration time step
of the Hamiltonian (\ref{arh}). We study
the temporal evolution of $\langle \Delta H_2(t_j)^2 \rangle$ over
the time span $t=10^6$. Fig.~\ref{var} presents the evolution of $H_2$'s  variance for the four smaller
values of $\kappa\in K$ where we observe that, within the regarded time interval, the variances increase in
mean linearly as expected. Moreover, though the diffusion domains do not overlap, the 
general properties of the chaotic dynamics provided by $\langle \Delta H_2^2\rangle$
seems to be quite similar since the variances take nearly the same values.  
However some departures from the linear trend are noticeable, in particular for $\kappa = 0.62$, which 
are due precisely to phase correlations. The variances for the remainder 
values of $\kappa$, which are not regarded for the figure, require a much longer time span for any 
increasing behavior to show up.

Let us now consider larger values of the parameters, namely, $\varepsilon = 0.25$ and $\mu = 0.1\varepsilon$, the
ones adopted in \cite{CGMB17}. In this case the approximation for the relative amplitude $v$ 
given in (\ref{v}) presents a  
maximum $v\approx 1.27$ at  $\kappa \approx 0.36$.
Thus we consider the following values of  $\kappa\in K'$:
$$K' = \{\kappa: \kappa = 0.25,\, 0.57,\, 0.75,\, 1.00,\, 1.50,\, 2.00,\, 9.00\},$$
such that $v$ ranges from $3\times 10^{-18}$ for the larger value of $\kappa$ up to
$1.22$ for the smallest one. For the remainder values of $\kappa = 0.57, 0.75, 1.00, 1.50, 2$
the relative amplitude amounts $1.02, 0.66, 0.31, 4.5\times 10^{-2}$ and $5\times 10^{-3}$ respectively.
We observe that for all values of $\kappa$,  $v\sim 1$ or smaller and thus  $\tau$ should exhibit strong 
correlations always
while it is expected that
for the largest values of $\kappa$, $\theta_2$ behaves nearly as random, accordingly to the discussion
given from the map (\ref{arm}). However, as it is shown in \cite{CGMB17}, since both $\varepsilon$
and $\mu$ are comparatively large, resonance interaction is strong leading to large variations
of $I_2$,  $-2\lesssim I_2\lesssim 2$ for all values of $\kappa$
(except perhaps for the largest one, see below) and thus the approximations introduced in the derivation of the map
seem to be no longer valid, i.e., $\omega_2$ could not be regarded as constant and moreover, the motion may be trapped 
in the resonance $\omega_2 = 0$. Additionally, the resonances $\omega_1=0$ and $\omega_1=\pm 1$ are close to
be in overlap and the amplitude of the resonances $\omega_1\pm\omega_2=0$ are not negligible, thus around
$\omega_2=1$ the motion could not be confined to the chaotic layer of the main resonance, as it was
discussed in \cite{CGMB17}. 
\begin{figure}[h!]
\sidecaption
\includegraphics[scale=0.6]{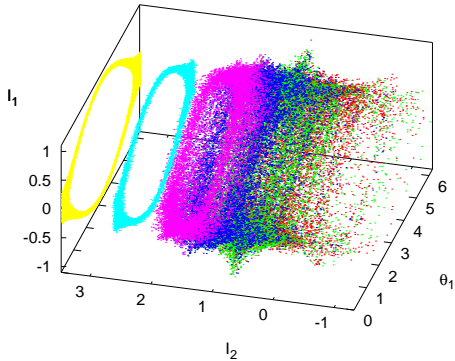}
\caption{Visualization of the diffusion along the chaotic layer of the resonance $\omega_1=0$ for 
	$\varepsilon = 0.25$ and $\mu = 0.1\varepsilon$ and
	six values of $\kappa\in K'$, $0.25$ in red, $0.57$ in green, $0.75$ in blue, $1.00$ in magenta,
$1.50$ in cyan and $2.00$ in yellow.
}
\label{dif3d025}
\end{figure}

Fig.~\ref{dif3d025} presents the visualization of the diffusion for the adopted values of $\varepsilon$ and $\mu$ and
several initial locations of the ensembles given by $\kappa\in K'$. 
We observe that for $0.25\le\kappa\le 1$ the diffusion domains overlap
and the action has a significant variation;  $-2\lesssim I_2\lesssim 2$, while for $\kappa > 1$,
the diffusion in $I_2$ is much more confined, similar to what is observed in Fig.~\ref{dif3d}. 
Notice that near $I_2 = 1$  diffusion spreads out of the chaotic
layer of the resonance $\omega_1 = 0$ where the motion is under the influence of at least three resonances,
$\omega_1 = 0$, $\omega_1 \pm \omega_2=0$ and $\omega_1 = \pm 1$. Thus, the numerical results confirm that 
for $\omega_2^0\lesssim 2$ the map (\ref{arm})
does no longer describe the diffusion along the chaotic layer of the main resonance.

Therefore for these large values of the perturbation parameters we numerically investigate the Hamiltonian
(\ref{arh}) instead of the map (\ref{arm}). To this aim, we take an ensemble of $n_p = 100$
random initial conditions on the separatrix of the resonance $\omega_1 = 0$. We integrate
the equations of motion up to $t = 5\times 10^5$ with a small time step $\delta t = 5\times 10^{-4}$ and
take the values of $\theta_2$ and $t\equiv\tau$ each time $|\theta_1 - \pi|\le 0.001$, avoiding two
consecutive crossings of the surface $\theta_1=\pi$ by a single trajectory\footnote{This could occur due to the
smallness of the time step.}. After normalizing
$\theta_2$ and $\tau$ to the unit interval, on adopting a sample time interval $\Delta t = 500$
and setting $q = 10^3$, we compute the information (\ref{info}) for the seven values of 
$\omega_2^0 = \kappa\sqrt{3}$, with $\kappa\in K'$. 

For the theoretical estimates,
the total number of points is now $N =n_p \times t/\delta t$, but since we are considering only those satisfying 
$|\theta_1 - \pi|\le 0.001$, we set $N = \beta n_p \times t/\delta t$, with $\beta\ll 1$, and therefore
(\ref{I}) and (\ref{Intorus}) become
\begin{equation}
	\mathcal{I}(\gamma_x^r,\alpha) \approx \frac{q\delta t}{2\beta n_p\ln q}\frac{1}{t},\qquad 
	\mathcal{I}(\gamma_x^s,\alpha) \approx \frac{q^2\delta t^2}{8\beta^2 \ln q}\frac{1}{t^2},
\label{IH}
\end{equation}
respectively, where $\beta \approx 3\times 10^{-4}$ obtained numerically.

\begin{figure}[t!]
\begin{tabular}{cc}
\hspace{-4mm}\includegraphics[scale=0.50]{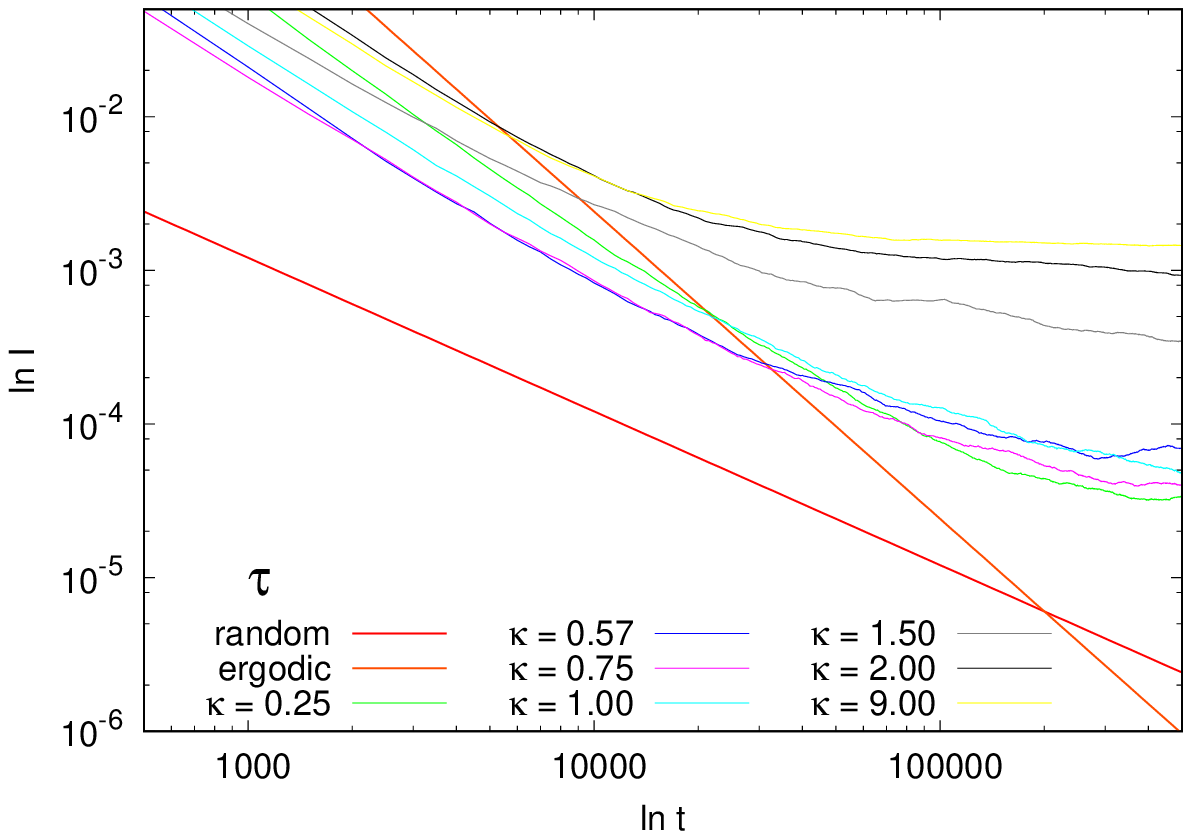}
\hspace{-2mm}\includegraphics[scale=0.50]{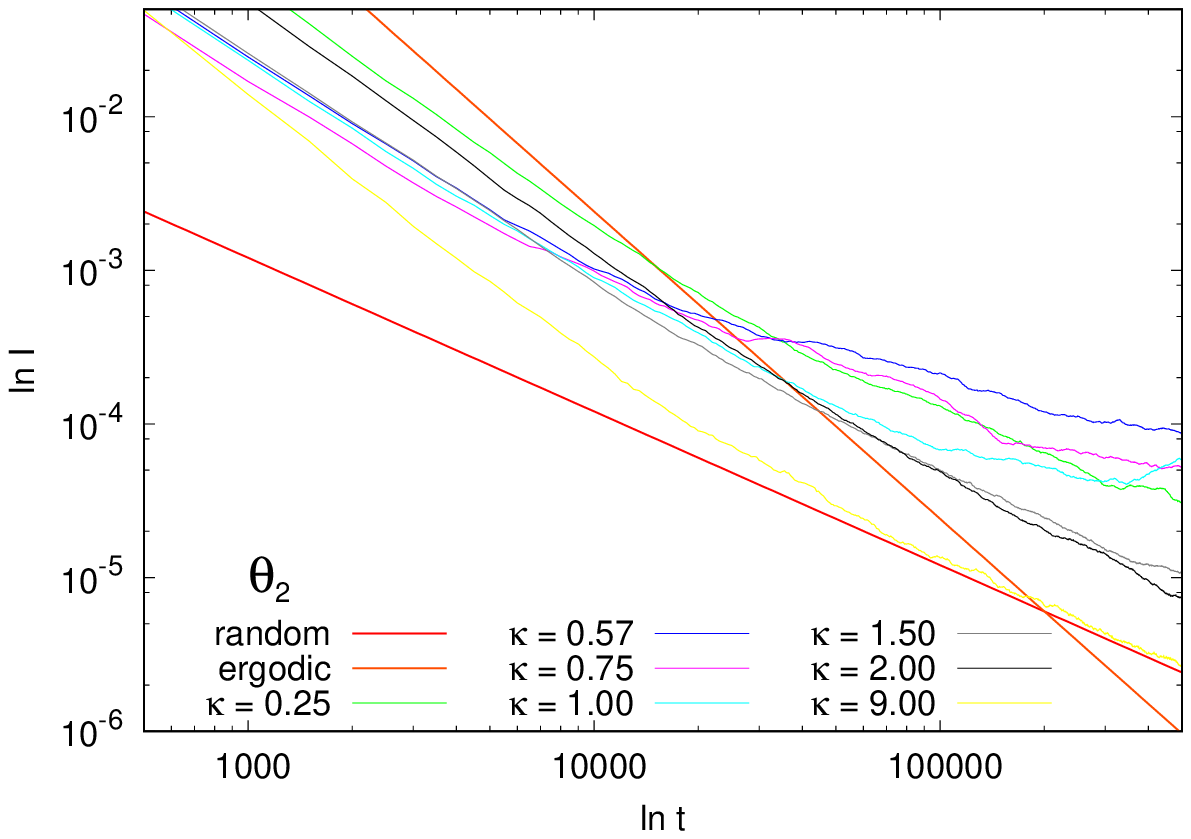}
\end{tabular}
        \caption{(Left panel) Evolution of $\mathcal{I}$ for the phase $\tau\in(0,1)$ corresponding to the values of $t$
        when the the pendulum crosses the surface $\theta_1=\pi$ for $n_p=100$ random initial values
        on the separatrix
        of the resonance $\omega_1 = 0$
        for the Hamiltonian (\ref{arh}) with $\varepsilon = 0.25, \mu = 0.1\varepsilon$ and seven
        values of $\omega_2^0 = \kappa\sqrt{3}$.
        The analytical expected behavior for random and ergodic motion  given by (\ref{IH}) are
        also included for $N = \beta n_p\times t/\delta t$, with $\beta \approx 3\times 10^{-4}$.
        (Right panel) Similar to the plot at the left but for
        the phase $\theta_2\in(0,1)$.}
\label{arnold2}
\end{figure}

Fig.~\ref{arnold2} shows the results for $\mathcal{I}$ for all the ensembles defined by $\omega_2^0 = \sqrt{3}\kappa$,
with $\kappa\in K'$. It becomes  evident that in any case, the phase $\tau$ is always correlated, though up to
$t\approx 10^4$ $\mathcal{I}$ decreases with a power law near to $t^{-1}$, at larger times 
decreases in a much slower way and for large $\kappa$ it reaches a nearly constant value at 
relatively short times. On the other hand, while $\theta_2$ decreases with time for all ensembles,
it could be assumed nearly random only for large values of $\omega_2$, precisely when 
$\omega_2\approx \mathrm{const}$ is a rough approximation. We observe that up to 
$t\approx 2\times 10^4$ it is $\mathcal{I}\sim t^{p}$ with $-2 < p < -1$,
but for larger times only those ensembles located at large $\omega_2^0$, $\mathcal{I}\sim t^{-1}$. Note
that for $\kappa = 9.00$, $\mathcal{I}$ behaves initially in an ergodic rather than random fashion, as expected
since $\theta_2\approx \omega_2 t$. 

In \cite{CGMB17} as well as in \cite{GC18} it was shown that for $\varepsilon = 0.25$ and
$\mu = 0.1\varepsilon$, the evolution of the ensemble variance (\ref{ea})
for $|\omega_2^0| < 2$ {reveals that the diffusion is clearly anomalous, in particular  a sub-diffusive process}. 
This fact motivated 
our introducing  an alternative way of measuring both the extent and the rate of the diffusion in \cite{GC18}, 
namely by means of the Shannon entropy.

Regarding the diffusion observed in $I_2$, we can say that for $\omega_2\lesssim 1$, the diffusion 
is certainly driven by all
first order resonances and it has no sense to speak about layer or driving resonances, all of them contributing 
to the motion along and across the chaotic layer of the guiding resonance. On the other hand, for $\omega_2\gg 1$
Chirikov's description could apply, resonance interaction is weak so the assumptions beneath the map (\ref{arm})
become plausible, the amplitude being $v\ll 1$ and regarding the evolution of $\theta_2$ for large $\omega_2$
we could take $\langle \cos^2\theta_2\rangle\approx R t/2$ with $R\lesssim 1$. 

Let us mention that in case of a general multidimensional Hamiltonian, Chirikov derived a similar map
than (\ref{arm}) but involving several phases and the discussion about the behavior of the different
phases is quite similar to the one given above (see \cite{Ch79}, \cite{C02}, \cite{CEGM14}).

\section{Discussion}\label{discussion}

Herein we have shown that the Shannon entropy turns out to be a rather simple and effective way to measure 
phase correlations. Even though we apply this technique to area--preserving maps or near--integrable Hamiltonian
systems, its formulation --given in Section~\ref{theory}-- reveals that it can be used in a wide range of problems
beyond dynamical systems. The numerical experiments included therein unfold that
the derived analytical estimates for absence of correlations or ergodicity are quite accurate.

The numerical simulations presented in this work concern in general chaotic motion, thus a given phase variable
cover completely the unit interval. Therefore, the key point in the present approach is the probability
of occupation of each element of the partition, $\mu(a_k)=n_k/N$. If the $n_k$ are randomly distributed, thus
the information $\mathcal{I}$ decreases as the inverse of the integer time or total number of phase values. 
On the other hand, if the $n_k$ have a uniform distribution, then the information decreases faster, following an inverse square law.  Any other dependence of $\mathcal{I}$ with $N$ reveals the existence
of correlations among the phase values.  Therefore a reference level for the information 
could be adopted for a given distribution of $N$ values of the phases, in order to discriminate 
between strong and weak correlations in the sample.

In dynamical systems like those studied in Section~\ref{experiments}, it would be much more interesting
to study how the information evolves with time. In such a way it is possible to reveal and understand
the underlying dynamics, as for instance in Figs.~\ref{standard} and \ref{whisker}.

More interesting is the use of this novel tool to revisit Chirikov's approach to diffusion along
the chaotic layer of a single resonance or Arnold diffusion, in a broad sense of this term.  We
succeed in revealing strong and weak correlations that prevent the free diffusion and therefore
the so--called reduced stochasticity approximation, in the sense that
$\langle \cos^2\theta_2\rangle\approx R t/2$ or $\langle \sin^2\tau\rangle\approx R t/2$ with $R < 1$
is not applicable and thus the obtained estimates for the diffusion coefficient are not correct 
(see \cite{Ch79}--Section 7, \cite{CGMB17} for more details).
Particularly, the map (\ref{arm}) could be a reliable model to investigate diffusion along
the chaotic layer of resonance $\omega_1 = 0$ provided that $\mu\varepsilon\ll\varepsilon\ll 1$
and $\omega_2$ is far away from $0$ and $1$ as Fig.~\ref{arnold1} shows. On the other hand,
if the parameters are not too small, the map would provide a fair approximation to
the diffusion that actually takes place only if $\omega_2\gg 1$. In other words, Chirikov's
formulation seems to be adequate when the diffusion extent is quite restricted and thus it
makes sense to derive a local diffusion coefficient.


\begin{acknowledgements}
This work was supported by grants from Consejo Nacional de Investigaciones Cient\'{\i}ficas y
T\'ecnicas de la Rep\'ublica Argentina (CONICET), the Universidad Nacional de La Plata 
and {Instituto de Astrof\'{\i}sica de La Plata}. We acknowledge two anonymous reviewers for
their valuable comments and suggestions that allow us to improve this manuscript. 
\end{acknowledgements}

\section*{Conflict of interest}
The authors declare that they have no conflict of interest.

\end{document}